%% file: Main.tex
\documentclass[conference,a4paper]{IEEEtran}
\usepackage[utf8]{inputenc}
\usepackage{amsmath}
\usepackage{amsfonts}
\usepackage{amssymb}
\usepackage{graphicx}
\usepackage{xcolor}
\usepackage{soul}
\usepackage{comment}
\usepackage{caption}
\usepackage{subcaption}

\usepackage{epsfig}  
\usepackage{multirow}
\usepackage{psfrag}
\usepackage[linesnumbered,ruled]{algorithm2e}
\usepackage{float}
\usepackage{diagbox}
\usepackage{url}
\usepackage{hyperref}
\usepackage{xcolor}
\usepackage{stfloats}
\usepackage{cite}
\usepackage[normalem]{ulem}

\usepackage{tabulary}

\definecolor{boristext}{rgb}{0.22, 0.44, 0.88}
\definecolor{boriscomments}{rgb}{0.88, 0.04, 0.04}
\definecolor{boristochange}{rgb}{0.2, 0.8, 0.8}

\title{Performance and Coexistence Evaluation of\\IEEE 802.11be Multi-link Operation}

\author{
\IEEEauthorblockN{Marc Carrascosa-Zamacois$^{\star}$, Lorenzo Galati-Giordano$^{\flat}$, Anders Jonsson$^{\star}$,\\Giovanni Geraci$^{\star}$, and Boris Bellalta$^{\star}$\vspace{0.2cm}
}
\IEEEauthorblockA{$^{\star}$\emph{Universitat Pompeu Fabra, Barcelona}}
\IEEEauthorblockA{$^{\flat}$\emph{Nokia Bell Labs, Stuttgart}}
\thanks{M. Carrascosa and B. Bellalta were supported in part by grants WINDMAL PGC2018-099959-B-I00 (MCIU/AEI/FEDER,UE), PRE2019-088690 (MCIU/FPI), and Cisco.
G. Geraci was in part supported by MINECO's Project RTI2018-101040 and by a ``Ram\'{o}n y Cajal" Fellowship from the Spanish State Research Agency.}
}

\IEEEoverridecommandlockouts
\begin{document}

\bstctlcite{IEEEexample:BSTcontrol}

\maketitle

\input{00_Abstract}
\input{01_Intro}

\input{02_MLO}
\input{03_Model}

\input{04_Results}

\input{05_Conclusion}

	\bibliography{Main.bbl}
	\bibliographystyle{IEEEtran}
\end{document}

%% file: 00_Abstract.tex
\begin{abstract}
Wi-Fi 7 is already in the making, and Multi-Link Operation (MLO) is one of the main features proposed in its correspondent IEEE 802.11be amendment. MLO will allow devices to coordinate multiple radio interfaces to access separate channels through a single association, aiming for improved throughput, network delay, and overall spectrum reuse efficiency. In this work, we study three reference scenarios to evaluate the performance of the two main MLO implementations---Multi-Link Multi-Radio (MLMR) and Multi-Link Single-Radio (MLSR)---, the interplay between multiple nodes employing them, and their coexistence with legacy Single-Link devices. Importantly, our results reveal that the potential of MLMR is mainly unleashed in isolated deployments or under unloaded network conditions. Instead, in medium- to high-load scenarios, MLSR may prove more effective in reducing the latency while guaranteeing fairness with contending Single-Link nodes. 
\end{abstract}

%% file: 01_Intro.tex
\bstctlcite{IEEEexample:BSTcontrol}

\section{Introduction}

Wi-Fi is more popular than ever. There will be 628 million Wi-Fi hotspots by 2023, four times up from 2018, 11\% of which adopting Wi-Fi 6 and 6E \cite{Cisco2020,WiFiAlliance2022}. Meanwhile, a new generation of Wi-Fi---IEEE 802.11be, or Wi-Fi 7---is in the making, with technical discussions underway to determine the specific implementation of several disruptive new features \cite{yang2020survey,lopez2019ieee,garcia2021ieee,khorov2020current,deng2020ieee,adame2021time}. The new capabilities of IEEE 802.11be will include 320~MHz bandwidth channels, 16 spatial streams, hybrid automatic repeat request (HARQ), 
and multi-band/channel aggregation and operation \cite{80211beDraft}. 

This last feature, commonly known as Multi-Link Operation (MLO), refers to the joint use of multiple radio interfaces on a single device. Owing to its promised augmented throughput and reduced delay, MLO is arguably the new feature drawing the most attention from industry and academia alike \cite{yang2019ap, song2020performance,lopez2022multi,lopez2022dynamic,naik2021can,lacalle2021analysis,carrascosa2021experimental}. However, the performance of specific MLO implementations, the interplay between multiple devices implementing MLO, and the coexistence of MLO with legacy channel access schemes are all crucial issues that remain largely unexplored.

In this paper, we bridge the above gap and investigate the performance of two MLO implementations as well as their coexistence with other legacy devices. In particular, we conduct extensive experiments comparing three channel access mechanisms: $i)$ traditional single-link (SL) operations, where a device avails of a single radio interface; $ii)$ multi-link single radio (MLSR), where multiple radio interfaces are available but only one at a time can be opportunistically used; and $iii)$ multi-link multi radio (MLMR), where the multiple available radio interfaces can be used concurrently. Our study unfolds as follows:
\begin{itemize}
\item We begin by considering an isolated Basic Service Set (BSS) setting devoid of channel contention. In this case, MLSR---only accessing one interface at time---can merely reduce the backoff time, only yielding anecdotal delay gains over SL. MLMR does curb the worst delays by five-fold when availing of a second interface, though adding a third interface provides diminishing returns.
\item We then consider two MLO BSSs contending for channel access. Contending MLSR BSSs retain the same delay as contention-free SL BSSs, as they opportunistically react to the evolving channel occupancy. However, MLMR BSSs may surprisingly incur higher delays than those of SL and MLSR, since they sometimes starve one another.
\item We conclude by assessing the coexistence between a MLO BSS and two independent legacy SL BSSs. A MLMR BSS boosts its throughput at the expense of a nearly equivalent reduction for the two coexisting SL BSSs. Nonetheless, MLMR also allows its SL neighbors to achieve lower delays in all cases except when these are highly loaded. 
\end{itemize}

%% file: 02_MLO.tex
\section{A Primer on Multi-link Operation} \label{mlodescription}

\begin{figure*}[ht]
    \centering
    \includegraphics[width=0.95\textwidth]{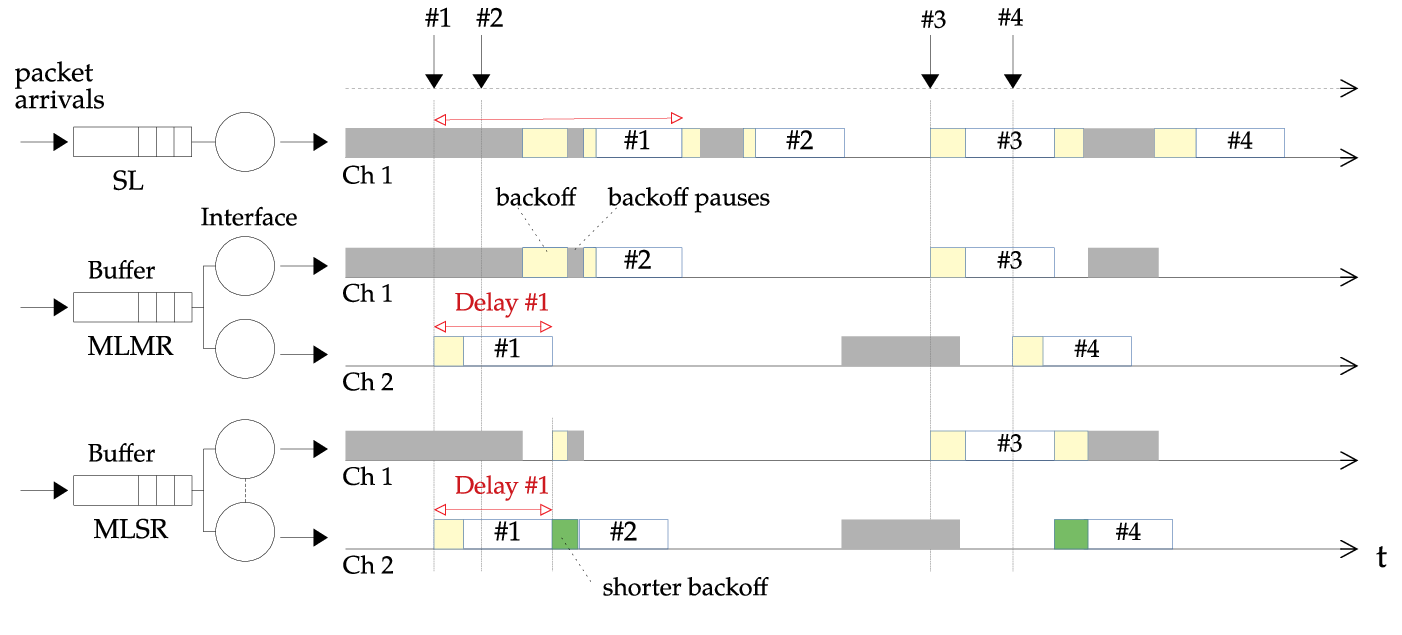}
    \caption{Illustration of SL, MLMR, and MLSR operations. The transmission representation (white box) includes control and acknowledgment frames.} 
    \label{Fig:MLO_howitworks}
\end{figure*}

MLO is being introduced in IEEE 802.11be to enable Wi-Fi devices to exchange data in a flexible manner over one or multiple wireless interfaces.\footnote{The terms link, channel, radio, and interface are used interchangeably.}
Compared to legacy SL devices, where multiple radios are operated through different and separate transmitter-to-receiver associations as if they were part of different BSS, MLO devices can benefit of using all available radios through a single association.   
Indeed, MLO devices can dynamically select one of the available interfaces, or even all at the same time, thus achieving opportunistic channel access or spectrum aggregation, respectively, and thus potentially higher transmission rates and lower delays. 

In this paper, we consider two MLO mechanisms that are likely to be found in the upcoming Wi-Fi 7 certified products (based on IEEE 802.11be). These two mechanisms, along with the legacy SL approach, are introduced in the sequel.
\begin{itemize}
	\item \textbf{Single-Link (SL)}: Default channel access, following the Distributed Coordination Function (DCF) and running over a single radio interface. At the transmitter side, when a packet is available in the transmission buffer, a backoff instance is initiated and the packet is transmitted once the backoff expires.
	\item \textbf{Multi-link Single Radio (MLSR)}: To support opportunistic spectrum access at a reduced cost, Wi-Fi devices can be equipped with a single fully functional 802.11be radio plus several other low-capability radios able only to decode IEEE 802.11 control packets (e.g., Wi-Fi preambles). 
	At the transmitter side, once a packet is available in the buffer, an independent backoff instance is initiated on each wireless interface, with the data packets then being allocated to only one of such interfaces according to a specific strategy, e.g., to the one whose backoff expires first. No other transmission is initiated until the one ongoing on the selected interface is completed.
	Once the transmitter determines the interface to use for the ongoing transmission, it informs the receiver, which in turn switches its fully functional 802.11be radio to the selected interface, receives the data packet, and responds with the corresponding ACK.
	\item \textbf{Multi-Link Multiple Radio (MLMR)}: For a device implementing this approach, all multiple radio interfaces are 802.11be compliant and they are able to operate concurrently, thus performing multiple simultaneous transmissions. 
	At the transmitter side, once a packet is available in the transmission buffer, a backoff instance is initiated on all inactive wireless interfaces, 
	allocating the data packets progressively to the interfaces as their backoffs expire.
    At the receiver side, packets are then received on all links used by the transmitter.
\end{itemize}
\subsubsection*{{Complexity}}
{Implementing the above three mechanisms (namely SL, MLSR, and MLMR) entails an increasing level of complexity. Indeed, SL employs a single 802.11 radio; MLSR requires an 802.11be radio as well as $S-1$ 'dummy' radios---$S$ being the number of interfaces---for channel sensing; MLMR requires $S$ full-blown 802.11be radios. We will show that the gains (or lack thereof) arising from an increased complexity may heavily depend on the specific scenario.}

\subsubsection*{{Example}}
Fig.~\ref{Fig:MLO_howitworks} exemplifies the operation of SL, MLSR, and MLMR. All the available interfaces (circles) share a single buffer and packets can thus be scheduled to either available interface. The figure illustrates the following:
\begin{itemize}
\item For SL, as packets arrive, the backoff starts and they are sent through the only available interface.
\item For MLMR, Packet 1 arrives while Channel 1 is busy and Channel 2 is idle, thus it is transmitted through the latter. During this transmission, another backoff instance begins on Channel 1 for Packet 2, resulting in transmitting both packets with shorter delays than in SL.
\item For MLSR, Packet 1 is also transmitted through Channel 2, and then the backoff is restarted on both channels. Packet 2 is sent through Channel 2, whose backoff expires first, and the backoff on Channel 1 is cancelled. 
\item As for Packets 3 and 4, MLMR transmits them simultaneously as soon as they arrive, with the transmission for Packet 4 starting during the transmission of Packet 3. 
\item For MLSR instead, Packet 3 is transmitted first on Channel 1. As Packet 4 arrives, it must wait for the ongoing transmission to be completed. Channel 2 then becomes available first and is used for Packet 4. Note that MLSR transmits Packet 4 more slowly than MLMR, but faster than SL.
\end{itemize}

%% file: 03_Model.tex
\section{Evaluation Methodology}
\label{model}

\begin{figure}[t]
\centering
    \includegraphics[width = \columnwidth]{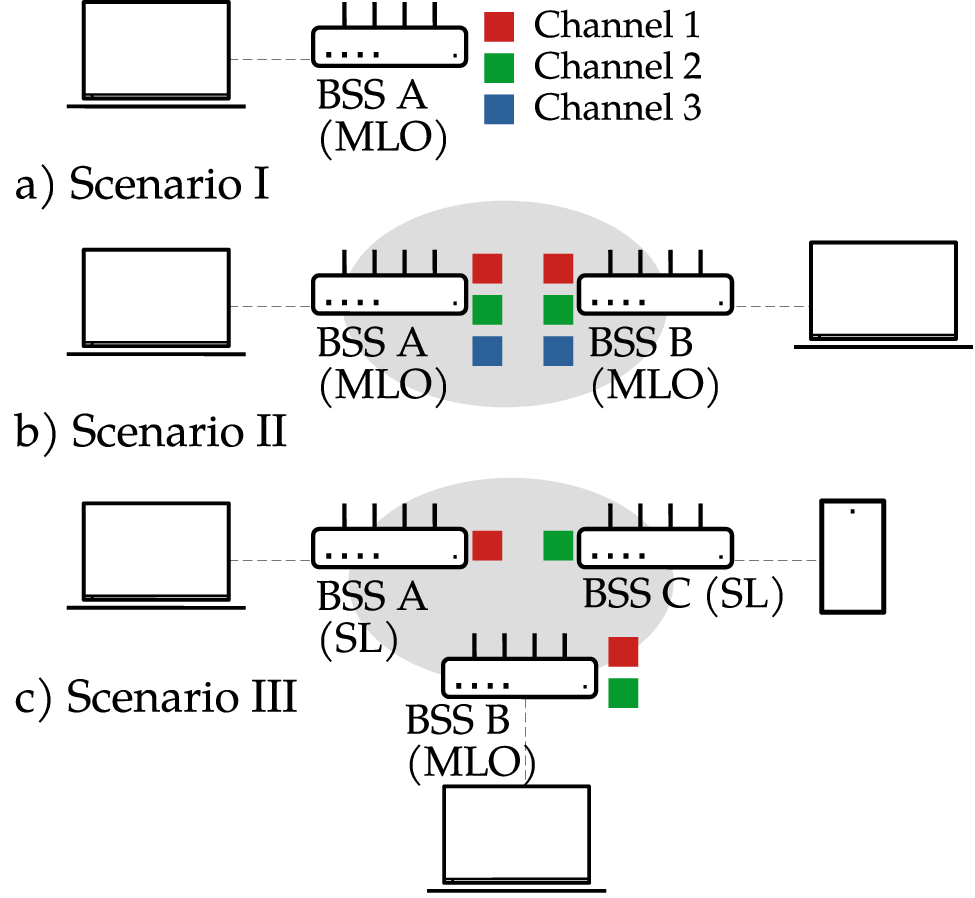}

    \label{Fig:scenario}
    \caption{Deployment scenarios considered in this paper: (a) a single MLO BSS, (b) two contending MLO BSSs, and (c) one MLO BSS coexisting with two legacy SL BSSs.} 
\label{Fig:scenario2}
\end{figure}

To carefully evaluate the performance of MLO as well as its coexistence with legacy SL, we consider the three representative scenarios depicted in Fig.~\ref{Fig:scenario2}. In particular:
\begin{itemize}
    \item \emph{Scenario I} is used to asses the performance gains experienced by a single and isolated BSS (BSS A) as it gets upgraded from SL to MLO.
    \item \emph{Scenario II} models the mutual interaction between two MLO BSSs (BSSs A and B) to study its effect on both.
    \item \emph{Scenario III} features a single MLO BSS (BSS B) and two independent SL BSSs (BSSs A and C), serving the all-important purpose of evaluating the coexistence of 802.11be MLO with legacy SL devices.
\end{itemize}

All three scenarios share the following features: $i)$ All BSSs are within each other's coverage area and therefore neither hidden terminal issues or deployment asymmetries arise; $ii)$ Only downlink traffic is considered, i.e., from the AP to an associated single station; $iii)$ Traffic arrival follows a Poisson process and all arriving packets have a constant size of $L=12000$ bits; $iv)$ APs have a transmission buffer size of 1000 packets; $v)$ A fixed modulation and coding scheme is employed, based on 256-QAM with rate 3/4 and 2 spatial streams\footnote{We set the transmission rate according to a link distance of $7$~m, a transmit power of 20~dBm, and the 802.11 TGax path loss model for residential scenarios \cite{tgax2018simulation}, resulting in a path loss of 72.51~dB and a received power of -58.51~dBm. No other channel impairments are considered, in order to isolate the effects of the three channel access schemes under consideration.}; $vi)$ A-MPDU packet aggregation is enabled for up to 64 packets, and the instantaneous number of aggregated MPDUs is chosen at the start of each transmission; $vii)$ The transmission duration depends on the A-MPDU size, thus ranging from 0.25 ms to 3.4 ms; $viii)$ The Request-to-Send/Clear-to-Send (RTS/CTS) mechanism is used to reserve the channel and, in the case of MLSR, to indicate which link will be used in the upcoming data transmission. The main system model parameters are summarized in Table
~\ref{table:parameters}. {In order to isolate the gains of MLO and its effect of legacy devices, the main system parameters are intentionally chosen according to 802.11be's predecessor and current standard, 802.11ax \cite{tgax2018simulation}.}

The MLO BSSs considered in the different scenarios are equipped with up to $3$ radio interfaces, each one operating on a different 80~MHz radio channel. 
The channel mapping strategy implemented by each BSS is depicted in Fig.~\ref{Fig:scenario2}, where the colored boxes denote the corresponding links/channels in use. In particular:
\begin{itemize}
    \item In Scenario I, BSS A uses Channels 1, 2, and 3.
    \item In Scenario II, both MLO BSSs use Channels 1, 2, and 3 simultaneously, thus modeling contention.
    \item In Scenario III, the two SL BSSs A and C employ orthogonal channels: Channel 1 and Channel 2, respectively. The MLO BSS B, employing both channels, thus contends with BSS A to access Channel 1 and with BSS C to access Channel 2. 
\end{itemize}
In all three scenarios, MLO BSSs operate according to either the MLSR or MLMR strategies presented in Section~\ref{mlodescription}. These new MLO features are implemented atop a Wi-Fi state machine originally developed to study channel bonding and spatial reuse under SL, thus bringing to the next level our previous work that only focused on IEEE 802.11ax networks \cite{bellalta2015interactions,wilhelmi2021spatial}. We carry out long-run simulations of 100~s, gathering traces with more than 150000 entries to guarantee an accurate characterization of both throughput and delay. \footnote{{While not reported for space constraints, the accuracy of our simulator was thoroughly validated using an extended version of the Markovian analytical models in \cite{bellalta2015interactions}, showing an excellent match.}}

\begin{table}
\centering
\caption{Wi-Fi state machine parameters for our studies.}
\label{table:parameters}
\def\arraystretch{1.2}
\begin{tabulary}{\columnwidth}{ |p{5.85cm} | p{2.1cm} | }
\hline
	\textbf{PHY} 			&  \\ \hline
	    Channel width & 80 MHz \\ \hline
		Modulation & 256 QAM 3/4 \\ \hline
		Transmission power & 20 dBm \\ \hline
		Legacy (HE single-user) preamble    & $20~\mu s$ ($52~\mu s$)\\ \hline
		OFDM (legacy) symbol duration & $16~\mu s$ ($4~\mu s$)\\ \hline
		Number of spatial streams & 2 \\ \hline\hline
	
	\textbf{MAC} 			&  \\ \hline
	    Short (DCF) InterFrame Space  & $16~\mu s$ ($34~\mu s$)\\ \hline
		Service field   & 32~bits\\ \hline
		MAC header  & 272~bits\\ \hline
		Tail (delimiter) bits & 6~bits (32~bits)\\  \hline
		ACK (block ACK) bits &  112~bits (256~bits)  \\ \hline 
		RTS (CTS) & 160~bits (112~bits) \\ \hline
		Frame size &  12000~bits\\ \hline
		A-MPDU size & 1--1024 packets\\ \hline
		Backoff (Best effort Access Category) &  CW$_{\min}$=15\\ \hline 
		AP buffer size & 4096 packets \\\hline
\end{tabulary}
\end{table}


%% file: 04_Results.tex
\section{Performance and Coexistence of MLO}

In this section, we consider the three scenarios described in Fig.~\ref{Fig:scenario2} and evaluate the performance of a MLO BSS as well as its coexistence with other MLO and legacy BSSs.\footnote{The values of throughput and delay ensue from the specific system model assumed. While the absolute values may differ from the performance limits of actual Wi-Fi 7 networks, their qualitative trends help understand the interplay and relative performance of the different approaches.}


\subsection{Scenario I: Single MLO BSS}\label{1bss}

We begin by evaluating performance gains provided by MLO in an isolated BSS setting, i.e., devoid of channel contention, as described in Scenario I. To this end, we compare the throughput and delay experienced by a MLO BSS to the one of a legacy SL BSS. We assume the latter to operate on a single radio interface, e.g., on Channel 1, and the former to jointly operate two or three radio interfaces, each on a different channel.

\begin{figure}
\centering
\begin{subfigure}[b]{\columnwidth}
    \includegraphics[width = \textwidth]{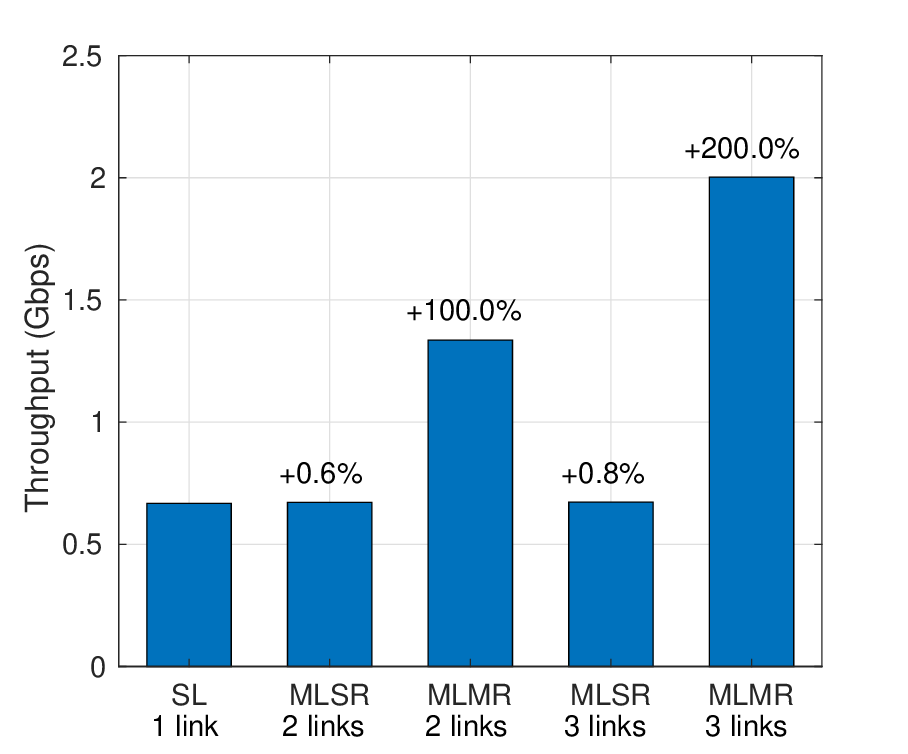}
\end{subfigure}
\caption{Scenario I: Throughput of a single BSS using Single-Link (SL) and Multi-Link (MLSR or MLMR) modes.}
\label{throughput1bss}
\end{figure}

Fig.~\ref{throughput1bss} shows the throughput achieved by each transmission method as a function of the number of radio interfaces. For MLSR, the throughput is almost identical to that of SL regardless of the number of available interfaces. Instead, the MLMR throughput increases linearly with the number of interfaces, i.e., two- and three-fold with two and three links, respectively.
Indeed, the latter is due to a lack of channel contention, allowing MLMR to transmit proportionally more data as the number of interfaces grows.

\begin{figure}
\centering
\begin{subfigure}[b]{\columnwidth}
    \includegraphics[width = \textwidth]{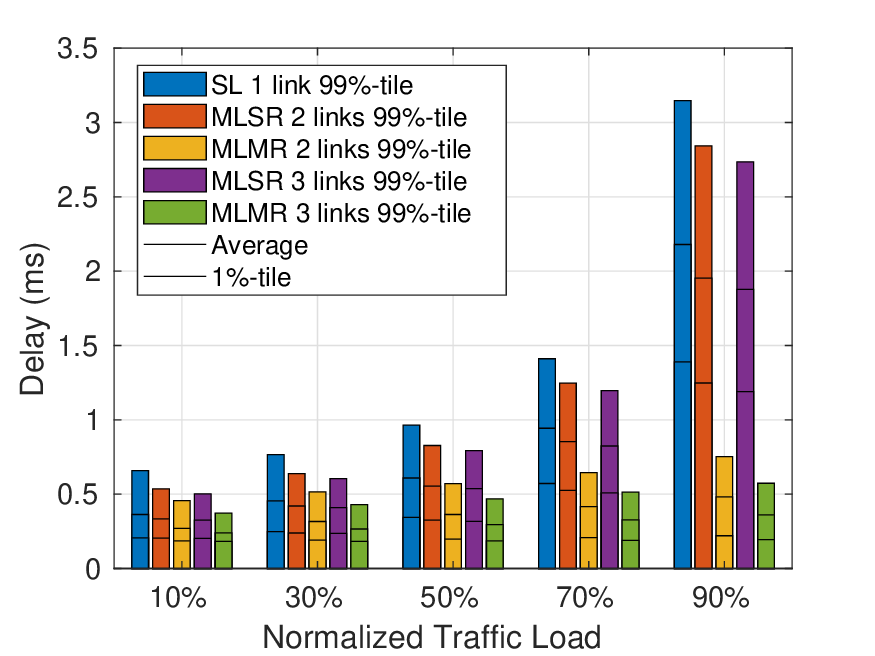}
\end{subfigure}
\caption{Scenario I: Delay of a single BSS using Single-Link (SL) and Multi-Link (MLSR or MLMR) modes. }
\label{delay1bss}
\end{figure}

Fig.~\ref{delay1bss} presents the associated average, 99\%-tile, and 1\%-tile delay, the last two serving as a proxy for worst- and best-case performance, respectively. The traffic load indicated is normalized to the SL throughput, i.e., 670 Mbps as per Fig.~\ref{throughput1bss}. For instance, 10\% and 30\% correspond to loads of 67 and 201~Mbps, respectively. 

As MLSR can only transmit through one radio interface at a time, its slightly reduced delay over SL is simply due to running simultaneous backoff counters and accessing the interface whose counter expires first. In the absence of channel contention, merely reducing the backoff time yields anecdotal delay reductions, as this is negligible with respect to the transmission time.

Unlike MLSR, MLMR does significantly decrease the delay since transmitting over multiple radio interfaces allows data packets to be received faster. MLMR is particularly effective at high loads, where availing of a second interface curbs the 99\%-tile delay at least by a factor of four. Note that, for the traffic loads considered, adding a third interface provides diminishing returns in delay reduction.

\subsection{Scenario II: Two Contending MLO BSSs}

\begin{figure}
\centering
\begin{subfigure}[b]{0.9\columnwidth}
    \includegraphics[width = \textwidth]{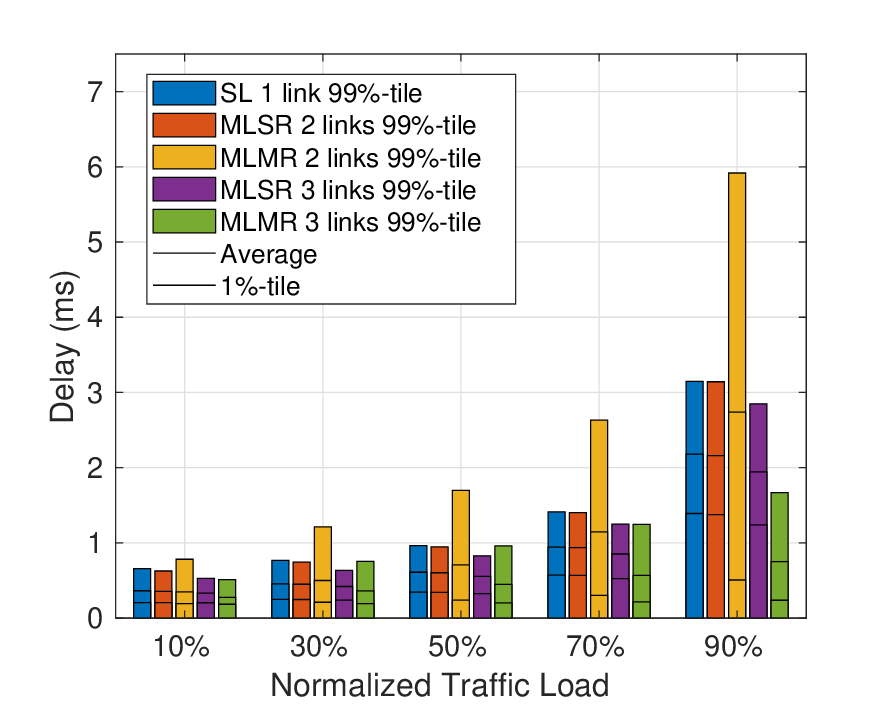}
\end{subfigure}
\caption{Scenario II: Delay performance for each transmission method as traffic load increases. Traffic load values refer to the fraction of the SL full-buffer throughput in Section \ref{1bss}. }
\label{delay}
\end{figure}

We now study the coexistence of two MLO BSSs contending for channel access, as described in Scenario II. The MLO BSSs are equipped with two or three radio interfaces each and can be operated either in MLSR or MLMR mode. Fig.~\ref{delay} shows the delay vs. traffic load under this setup as compared to the one experienced by two SL BSSs operating on orthogonal channels (and thus not contending for access).\footnote{While not shown for brevity, SL, MLSR, and MLMR all attain the same throughput as they handle the same incoming traffic load which is all successfully delivered.}

On the one hand, contending MLSR BSSs---with either two or three interfaces each---retain the same delay as contention-free SL BSSs. The opportunistic use of a single radio allows MLSR to react to the evolving contention levels over different channels. {Additionally, since MLSR with two interfaces always leaves at least one channel idle to the contending BSS, it results in a fairer share of the spectrum. In the case of MLSR with three interfaces, an extra backoff instance can be allocated to each MLO BSS, further reducing the channel access delay.}

On the other hand---and despite its higher complexity---MLMR {with two interfaces} somewhat surprisingly incurs higher 99\%-tile delays than those of SL and MLSR, even at loads as low as 30\%. Since MLMR BSSs can transmit through multiple interfaces at once, they can sometimes starve one another. As a result of this greedy policy, the best-case delays (e.g., 1\%-tile) are reduced but the worst-case ones (e.g., 99\%-tile) are increased. {A workaround to this shortcoming is to add an extra interface to each MLO BSS. Although the channel used by the extra interface is to be shared between the two MLO BSSs, its presence significantly increases the likelihood of finding an idle interface. Indeed, the 99\%-tile delay with a third interface drops to around half that of SL.}

\subsection{Scenario III: Contending MLO and SL BSSs}

We conclude by assessing the coexistence between a MLO BSS and two independent legacy SL BSSs, as in Scenario III. 
Specifically, we assume two radio interfaces and a constant traffic load for the MLO BSS (BSS B), and consider two different traffic loads for the SL BSSs (BSS A and BSS C), namely symmetric and asymmetric. Our aim is to shed light on how MLSR/MLMR affect the performance of neighboring legacy BSSs and how MLO BSSs handle symmetric and asymmetric activity in their operating channels.

\begin{figure}
    \includegraphics[width=0.9\columnwidth]{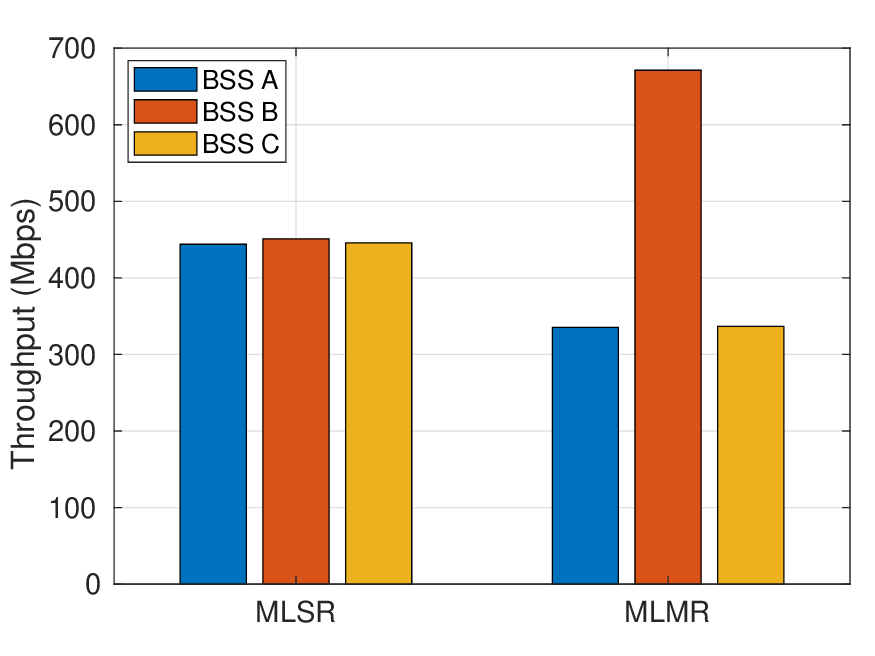}
    \caption{Scenario III: individual BSS throughput when BSS B employs MLSR (left) and MLMR (right). BSSs A and C are assume to employ a single link.} \label{delay3-2}
\end{figure}

Fig.~\ref{delay3-2} shows the full-buffer throughput achieved by all three BSSs when BSS B employs either MLSR or MLMR. 
When operating in MLSR mode, BSS B achieves almost identical throughput as SL BSSs A and C. However, when employing MLMR, BSS B boosts its own throughput at the expense of a nearly equivalent reduction for the two coexisting SL BSSs A and C.

\begin{figure}
\centering
\begin{subfigure}[b]{0.9\columnwidth}
    \includegraphics[width = \textwidth]{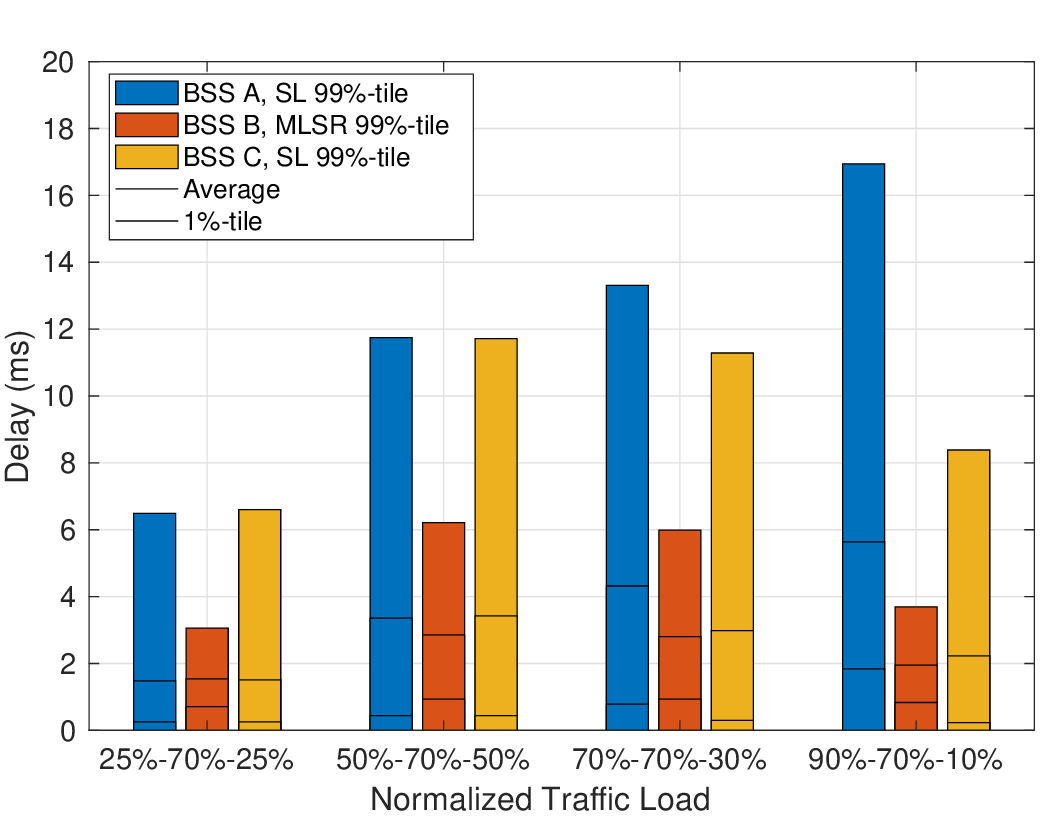}
    \caption{MLSR delay}
    \label{delay3-3}
\end{subfigure}
\begin{subfigure}[b]{0.9\columnwidth}
    \includegraphics[width = \textwidth]{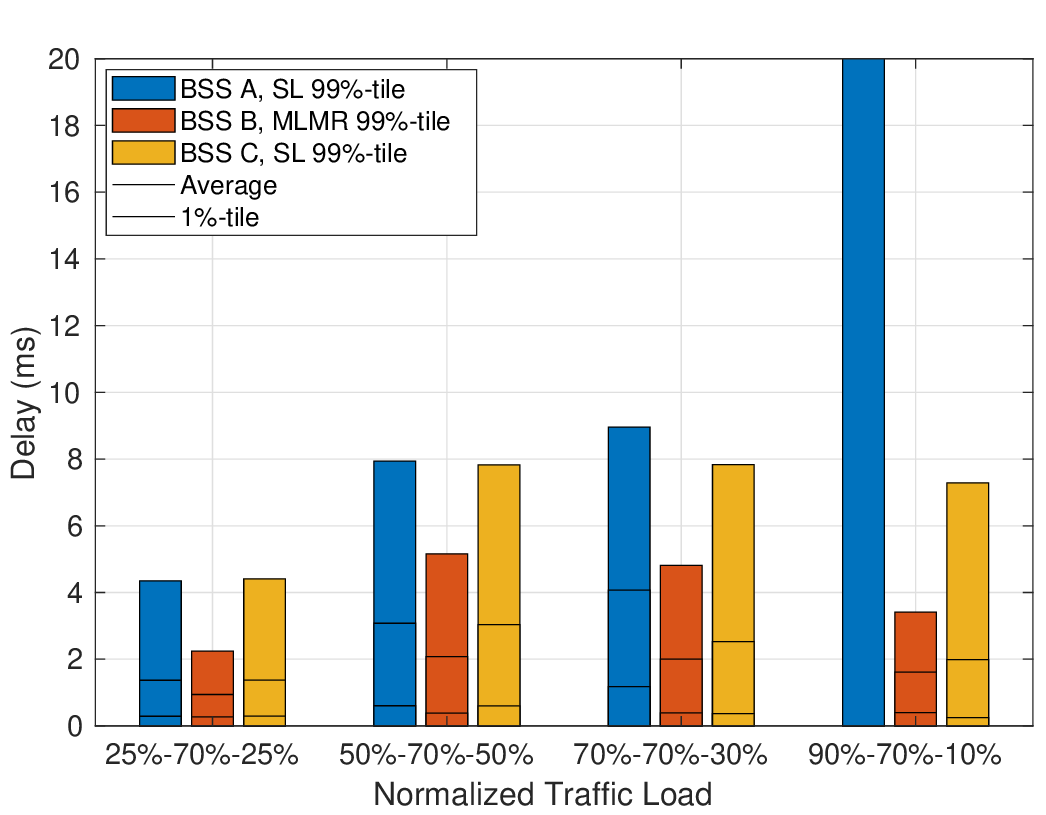}
    \caption{MLMR delay}
    \label{delay3-4}
\end{subfigure}
\caption{Delay in unbalanced scenarios. MLSR has a low impact on nearby BSSs, while MLMR has a severe impact on highly loaded channels, saturating BSS A. Traffic load values refer to the fraction of the SL full-buffer throughput obtained in Section \ref{1bss}.}
\label{delay3bss}
\end{figure}

Fig.~\ref{delay3-3} and Fig.~\ref{delay3-4} show the delay for each BSS when BSS B employs MLSR and MLMR, respectively. We consider several combinations for the traffic load fed to each BSS, indicated again as a fraction of the full-buffer throughput achieved by a SL BSS in Scenario I (670~Mbps as per Fig.~\ref{throughput1bss}). Specifically, the load on BSS B is set to 70\% while the loads on BSSs A and C are varied to model symmetric scenarios (with BSSs A and C experiencing the same load) and asymmetric scenarios (with BSS C increasingly less loaded than BSS A).

Both Fig.~\ref{delay3-3} and Fig.~\ref{delay3-4} show that MLSR/MLMR can opportunistically leverage the availability of an emptier channel. For instance, by comparing case \{50\%--70\%--50\%\} to case \{90\%--70\%--10\%\} with same aggregated contending traffic from BSSs A + C, we note that the latter results in a lower delay for MLSR/MLMR. In all traffic configurations considered, employing MLMR is only slightly more beneficial than MLSR for BSS B. As for the coexistence between the multi-link BSS B and the SL BSSs A and C, MLMR allows its SL neighbors to achieve lower delays than MLSR does in most cases. However when a SL BSS is highly loaded (i.e., 90\%), MLMR can occasionally cause it to starve and thus experience higher delays.

%% file: 05_Conclusion.tex
\section{Conclusion}	

This work is devoted to the understanding of the performance of two specific MLO implementations, the interplay between multiple devices implementing them, and their coexistence with legacy single-link channel access schemes. Through our extensive study---which compared $i)$ traditional single-link operations, $ii)$ multi-link single radio (MLSR), and $iii)$ multi-link multi radio (MLMR)---, we were able to draw the following key insights:
\begin{itemize}
\item In an isolated BSS setting devoid of channel contention,  MLMR with two interfaces can reduce the worst delays by a factor of five, whereas adding a third interface provides immaterial extra gains.
\item Two contending MLSR BSSs experience same delay as SL does in a contention-free setup. Surprisingly, {and despite its increased complexity}, MLMR BSSs may instead incur higher delays by occasionally preventing one another from timely accessing the channel.
\item When surrounded by legacy SL BSSs, a MLMR BSS boosts its own throughput at the expense of its SL neighbors', but also allows them to achieve lower delays for low-to-medium traffic loads.
\end{itemize}

{
The present work is, to the best of our knowledge, the first providing a well-grounded performance comparison of the two most relevant implementations of MLO and addressing the critical aspect of backward-compatibility with legacy SL devices.
Extensions are underway from different standpoints: 
\subsubsection*{Non-Poisson traffic}
By considering non-Poisson traffic with batch arrivals---a key feature, being MLO capable of transmitting multiple packets in the same batch at once.
\subsubsection*{Reproducibility}
By capturing the behavior of various MLO modes analytically, thus allowing a more generalized comparison and wide reproducibility of the results.
\subsubsection*{Interplay}
By studying the performance gains of MLO when paired with other new features being introduced in IEEE 802.11be and beyond, e.g., advanced AP coordination \cite{garcia2021ieee}.}

%% file: Main.bbl
\begin{thebibliography}{10}
\providecommand{\url}[1]{#1}
\csname url@samestyle\endcsname
\providecommand{\newblock}{\relax}
\providecommand{\bibinfo}[2]{#2}
\providecommand{\BIBentrySTDinterwordspacing}{\spaceskip=0pt\relax}
\providecommand{\BIBentryALTinterwordstretchfactor}{4}
\providecommand{\BIBentryALTinterwordspacing}{\spaceskip=\fontdimen2\font plus
\BIBentryALTinterwordstretchfactor\fontdimen3\font minus
  \fontdimen4\font\relax}
\providecommand{\BIBforeignlanguage}[2]{{%
\expandafter\ifx\csname l@#1\endcsname\relax
\typeout{** WARNING: IEEEtran.bst: No hyphenation pattern has been}%
\typeout{** loaded for the language `#1'. Using the pattern for}%
\typeout{** the default language instead.}%
\else
\language=\csname l@#1\endcsname
\fi
#2}}
\providecommand{\BIBdecl}{\relax}
\BIBdecl

\bibitem{Cisco2020}
{Cisco Annual Internet Report (2018--2023) White Paper}, March 2020, accessed
  on 05/08/2022.

\bibitem{WiFiAlliance2022}
{Wi-Fi alliance Wi-Fi 6E insights}, April 2022, accessed on 05/08/2022.

\bibitem{yang2020survey}
M.~Yang and B.~Li, ``{Survey and perspective on extremely high throughput (EHT)
  WLAN—IEEE 802.11 be},'' \emph{{Mobile Networks and Applications}}, vol.~25,
  no.~5, pp. 1765--1780, 2020.

\bibitem{lopez2019ieee}
D.~L{\'o}pez-P{\'e}rez, A.~Garcia-Rodriguez, L.~Galati-Giordano, M.~Kasslin,
  and K.~Doppler, ``{IEEE 802.11be Extremely High Throughput: The next
  generation of Wi-Fi technology beyond 802.11ax},'' \emph{{IEEE Commun.
  Mag.}}, vol.~57, no.~9, pp. 113--119, 2019.

\bibitem{garcia2021ieee}
A.~Garcia-Rodriguez, D.~L\'{o}pez-P\'{e}rez, L.~Galati-Giordano, and G.~Geraci,
  ``{IEEE 802.11be: Wi-Fi 7 strikes back},'' \emph{{IEEE Commun. Mag.}},
  vol.~59, no.~4, pp. 102--108, 2021.

\bibitem{khorov2020current}
E.~Khorov, I.~Levitsky, and I.~F. Akyildiz, ``{Current status and directions of
  IEEE 802.11be, the future Wi-Fi 7},'' \emph{{IEEE Access}}, vol.~8, pp.
  88\,664--88\,688, 2020.

\bibitem{deng2020ieee}
C.~Deng, X.~Fang, X.~Han, X.~Wang, L.~Yan, R.~He, Y.~Long, and Y.~Guo, ``{IEEE
  802.11be Wi-Fi 7: New challenges and opportunities},'' \emph{{IEEE Commun.
  Surveys \& Tuts.}}, vol.~22, no.~4, pp. 2136--2166, 2020.

\bibitem{adame2021time}
T.~Adame, M.~Carrascosa-Zamacois, and B.~Bellalta, ``{Time-sensitive networking
  in IEEE 802.11be: On the way to low-latency WiFi 7},'' \emph{{Sensors}},
  vol.~21, no.~15, p. 4954, 2021.

\bibitem{80211beDraft}
``{IEEE P802.11be/D1.5 - Draft Standard for Information technology--
  Telecommunications and information exchange between systems Local and
  metropolitan area networks-- Specific requirements - Part 11: Wireless LAN
  Medium Access Control (MAC) and Physical Layer (PHY) Specifications -
  Amendment 8: Enhancements for extremely high throughput (EHT)},'' March 2022.

\bibitem{yang2019ap}
M.~Yang, B.~Li, Z.~Yan, and Y.~Yan, ``{AP} coordination and full-duplex enabled
  multi-band operation for the next generation {WLAN}: {IEEE 802.11be (EHT)},''
  in \emph{{Proc. WCSP}}, 2019, pp. 1--7.

\bibitem{song2020performance}
T.~Song and T.~Kim, ``{Performance analysis of synchronous multi-radio
  multi-link MAC protocols in IEEE 802.11be extremely high throughput WLANs},''
  \emph{{Applied Sciences}}, vol.~11, no.~1, p. 317, 2020.

\bibitem{lopez2022multi}
{\'A}.~L{\'o}pez-Ravent{\'o}s and B.~Bellalta, ``Multi-link operation in {IEEE
  802.11be WLANs},'' \emph{IEEE Wireless Commun.}, pp. 1--12, 2022.

\bibitem{lopez2022dynamic}
{\'A}.~L{\'o}pez-Ravent{\'o}s and B.~Bellalta, ``Dynamic traffic allocation in
  {IEEE 802.11be} multi-link {WLANs},'' \emph{IEEE Wireless Commun. Letters},
  vol.~11, no.~7, pp. 1404--1408, 2022.

\bibitem{naik2021can}
G.~Naik, D.~Ogbe, and J.-M.~J. Park, ``{Can Wi-Fi 7 support real-time
  applications? On the impact of multi link aggregation on latency},'' in
  \emph{Proc. IEEE ICC}, 2021, pp. 1--6.

\bibitem{lacalle2021analysis}
G.~Lacalle, I.~Val, O.~Seijo, M.~Mendicute, D.~Cavalcanti, and
  J.~Perez-Ramirez, ``{Analysis of latency and reliability improvement with
  multi-link operation over 802.11},'' in \emph{Proc. IEEE INDIN}, 2021, pp.
  1--7.

\bibitem{carrascosa2021experimental}
M.~Carrascosa, G.~Geraci, E.~Knightly, and B.~Bellalta, ``An experimental study
  of latency for {IEEE} 802.11be multi-link operation,'' in \emph{Proc. IEEE
  ICC}, 2022, pp. 1--6.

\bibitem{tgax2018simulation}
{IEEE 802.11 TGax}, ``{TGax Simulation Scenarios},''
  \url{https://mentor.ieee.org/802.11/dcn/14/11-14-0980-14-00ax-simulationscenarios.docx},
  accessed on 05/08/2022.

\bibitem{bellalta2015interactions}
B.~Bellalta, A.~Checco, A.~Zocca, and J.~Barcelo, ``{On the interactions
  between multiple overlapping WLANs using channel bonding},'' \emph{IEEE
  Transactions on Vehicular Technology}, vol.~65, no.~2, pp. 796--812, 2015.

\bibitem{wilhelmi2021spatial}
F.~Wilhelmi, S.~Barrachina-Mu{\~n}oz, C.~Cano, I.~Selinis, and B.~Bellalta,
  ``{Spatial reuse in IEEE 802.11ax WLANs},'' \emph{Computer Communications},
  vol. 170, pp. 65--83, 2021.

\end{thebibliography}
